\documentclass[letterpaper, 10 pt, conference]{ieeeconf}
\usepackage{graphicx}      
\IEEEoverridecommandlockouts

\usepackage{cite}
\usepackage{amsmath,amssymb,amsfonts}
\usepackage{algorithmic}
\usepackage{graphicx}
\usepackage{textcomp}
\usepackage{xcolor}
\usepackage[htt]{hyphenat}
\usepackage{lipsum}
\def\BibTeX{{\rm B\kern-.05em{\sc i\kern-.025em b}\kern-.08em
    T\kern-.1667em\lower.7ex\hbox{E}\kern-.125emX}}
\usepackage{xspace}
\usepackage{soul}
\usepackage{mathtools}
\usepackage{makecell}
\usepackage{nicefrac}
\usepackage{float}
\usepackage{wrapfig}
\usepackage{amsmath}

\usepackage{cancel}

\newcommand{\colvec}[2][.8]{%
  \scalebox{#1}{%
    \renewcommand{\arraystretch}{.8}%
    $\begin{bmatrix}#2\end{bmatrix}$%
  }
}

\newcommand{\tra}[2]{\eta_{#1}^{#2}}

\definecolor{dgreen}{rgb}{0.0, 0.5, 0.0}

\makeatletter
\newcommand{\subalign}[1]{%
	\vcenter{%
		\Let@ \restore@math@cr \default@tag
		\baselineskip\fontdimen10 \scriptfont\tw@
		\advance\baselineskip\fontdimen12 \scriptfont\tw@
		\lineskip\thr@@\fontdimen8 \scriptfont\thr@@
		\lineskiplimit\lineskip
		\ialign{\hfil$\m@th\scriptstyle##$&$\m@th\scriptstyle{}##$\crcr
			#1\crcr
		}%
	}
}

\newcommand{\mct}[1]{{\fontfamily{cmtt}\selectfont #1}}
\newcommand{\mctm}[1]{\mbox{\fontfamily{cmtt}\selectfont #1}}

\newenvironment{myitemize}
{ \begin{itemize}
    \setlength{\itemsep}{0pt}
    \setlength{\parskip}{0pt}
    \setlength{\parsep}{0pt}     }
{ \end{itemize} }

\newif\if@secthm
\if@secthm
\newtheorem{thm}{Theorem}[section]
\@addtoreset{thm}{section}
\else
\newtheorem{thm}{Theorem}
\fi

\newtheorem{definition}[thm]{Definition}

\newtheorem{remark}[thm]{Remark}

\title {Formal verification of a controller implementation in fixed-point arithmetic} 

\author{Lars Flessing, Grigory Devadze, Stefan Streif$^*$
\thanks{$^*$ The authors are with the Automatic Control and System Dynamics Lab, Technische Universität Chemnitz, Chemnitz, Germany \{lars.flessing, grigory.devadze, stefan.streif\}@etit.tu-chemnitz.de.}}

\begin{document}
\maketitle
\thispagestyle{empty}
\pagestyle{empty}
\begin{abstract}                
  For the implementations of controllers on digital processors, certain limitations, e.g. in the instruction set and register length, need to be taken into account, especially for safety-critical applications.
  This work aims to provide a computer-certified inductive definition for the control functions that are implemented on such processors accompanied with the fixed-point data type in a proof assistant.
  Using these inductive definitions we formally ensure correct realization of the controllers on a digital processor. Our results guarantee overflow-free computations of the implemented control algorithm.
  The method presented in this paper currently supports functions that are defined as polynomials within an arbitrary fixed-point structure.
  We demonstrate the verification process in the case study on an example with different scenarios of fixed-point type implementations.
\end{abstract}

\section{Introduction}
Verification is an important part in the development of control systems.
Since some control systems are used in safety-critical applications, such as autonomous driving~\cite{Kong15AutoDirveCont} or avionics \cite{delmas2009towards,goodloe2013verification}, it is essential to ensure that they meet certain performance and safety requirements.
One key challenge is to guarantee that the properties of the control algorithm which are usually proven mathematically, are retained after numerical implementation in software~\cite{holzmann2001software}.
Furthermore, certain standards must be met by the final realization i.e. the software which runs on the digital processor. 

Conventionally, it is often assumed that the numerical imperfections are captured by the input error or system uncertainty which are then dealt within a robust control setting~\cite{Schulting19Num}.
However, in the case of an overflow within the calculations, the resulting values depend on the realization of the controller itself and may not be bounded \cite{yates2009fixed}.
In the so-called algorithmic control theory \cite{Tsiotras2017-toward}, the importance of dealing with these imperfections by suitable formal approaches has been highlighted.
The present work contributes toward these goals and aims to guarantee the correctness of the implementation of the controllers by bridging the gap between mathematical notion and the actual software implementation. This approach differs from the classical numerical analysis, where it is often unclear whether the used concepts are connected to the actual software code. In contrast to that, we derive numerical and computational concepts within a special so-called type-theoretic realization.
In particular, the present work relies on the use of the proof assistant software and the so-called \textit{proofs-as-programs paradigm}.

\section{Related works}
The main concern for certain safety-critical applications is the verification of performance characteristics for a given controller design.
In particular, such verification can be done using closed-loop simulation~\cite{kapinski2016simulation}.
This method provides the ability to test a predefined range of scenarios which are relevant for the behavior of the systems.
A common method for the verification purposes is the so-called \textit{fault injection} into the simulation system~\cite{Hsueh97FaultInj}.
Yang et al. use a fault injection strategy for traction control systems in the simulation of a high-speed trains~\cite{Yang17FaultInj}.
In~\cite{Yang18FaultITL}, the authors extend this approach to the field tests where a hardware-in-the-loop solution mimics hardware faults.

Another solution to ensure safety-critical characteristics of a control system is to add fail-safe systems and fault modes, which is known as \textit{fault-tolerant control}~\cite{JIANG201260}.
For example, the control function can be divided into different modes that handle different performance requirements.
These additional modes can be used to specifically ensure that the desired requirements are met.
As an example, in~\cite{Xiang16LTIFailSafe} the authors introduce an error mode for a linear time-invariant control system that suffers from a loss of control input.
The error mode is defined as a separate control function which has to ensure system stability. 
Blanke et al.~\cite{blanke2006diagnosis} presented a variety of approaches, e.g.~fault tolerant \(\mathcal{H}_{\infty}\)-design or fault tolerant model fitting designs.
Keel et al. discussed inaccuracy effects of linear robust controller.
The authors introduced the notion of fragility that is the loss of the controller performance caused by the deviation of the control parameters and defined respective metrics to estimate the fragility of a robust controller~\cite{RobustFragile}.

One may aim to analyze correctness and error-proneness via statistical approaches. In\cite{daza2020error}, the authors use uncertainty quantification~\cite{smith2013uncertainty} to make a qualitative statement about the computational correctness.
This approach has been used by Michelmore et al. to provide statistical guarantees for autonomous vehicle control~\cite{Michelmore20UQAutoDrive} or by Berning et al. for random constrained path planning of unmanned aircraft~\cite{berning2020rapid}.
However, these approaches are more of numerical nature and do not have a formal connection to the actual software realization, i.e. all approaches assume that the underlying controllers are implemented correctly.

To address the software implementation and obtain specific certification there is no way around of the usage of formal methods.  
Formal methods can be used to assert a guarantee for a certain mathematical property within a logical system and prove that this property holds.

The core of several tools is a \textit{satisfiability modulo theories} (SMT) solver~\cite{barrett2018satisfiability} that is able to check a set of first-order logical statement. Bessa et al. presents a framework that converts the source code of the digital controller with a processor definition into a set of logical instructions~\cite{VerificationFWLDigiContr}.
The tool \textit{RoSa} uses a combination of SMT solving and interval arithmetic to derive finite precision error bounds via verification constraints~\cite{RoSa}.

Herencia-Zapana et al. use source code annotations over Lyapunov-stabilizing properties of the code lines described via the quadratic invariants~\cite{PVSLinAlgForC}.
With these annotations, the authors obtain a translation of the code into the representation of the associated logical statements, which are checked within the prototype verification system PVS~\cite{owre1992pvs}.
Alternatively, Roozbehani et al. proposed a control-theoretic framework for the verification of numerical programs \cite{Roozbehani2013-Lyapunov}.
The authors rely on the framework of the so-called Lyapunov invariants and employ the convex optimization for their construction.
Recently, Devadze et al. used computer-assisted proofs and program extraction techniques for the formalization of initial value problem for ODE and sum-of-squares certificates for stability of polynomial systems \cite{devadze2020ODE,devadze2020computer}. An alternative direction of algorithmic verification is \textit{FLUCTUAT}, which allows to propagate the errors of function on variables with an uncertainty margin using statistical analysis.
Finally, we mention the framework  \textit{FPTaylor}, where roundoff errors are estimated by a symbolic Taylor expansion with the help of HOL~\cite{SymTaylorExp} and the library \textit{Real2Float}, where one is able to estimate an upper error bound for floating-point programs~\cite{Real2Float}.


\subsection{Contributions and outline of this work.}
The focus of the present work is to provide a formal framework for verified controller implementations in which computations are guaranteed to be overflow-free (\textit{OF}).
This is obtained by modeling the behavior of the processor hardware as a logical type of the used mathematical interpretation.
Thus we may derive a mathematical representation of the effects of the hardware definition, such as the overflow effect that is presented in this paper.
We extensively use formal methods and the proof-as-programs paradigm for the sake of correctness.
Section~\ref{sec::Problem} introduces the problem setup and the motivating example, where the occurrence of the overflows have destabilizing effects on system behavior.  
For demonstration purposes the example is carried out with the simulation of different fixed-point type settings to show the essence of the verification process.
Section~\ref{sec::Prelim} presents the proof assistant \mct{Minlog} and provides the formal definitions for the data types that model the processor arithmetic.
The derivation of the correct controller setup for the overflow-free computation is given in Section~\ref{sec::MainRes}.
In Section~\ref{sec::CaseS} we demonstrate the verification process by revisiting the example from the Section~\ref{sec::Problem} and show that the certain controller implementation, which led to the actual failure, did not satisfy the formal definition.
Finally, we provide the conclusion and an outlook in Section~\ref{sec::Outl}.




\section{Problem Setup}
\label{sec::Problem}
We begin with a motivational example.

\subsection{Controller implementation}
\label{sec::SysDef}
We consider the implementation of a controller on a digital processor (e.g. micro-controller) using fixed-point arithmetic. Since the digital processor provides only a limited number of bits for the representation, the implementation is affected by the so-called \textit{finite word length problem}.
For a fixed-point specification with \(q \in \mathbb{N}\) bits for the integer part and \(p\in\mathbb{N}\) bits for the fraction part the respective value domain of the controller  is defined by
\begin{align}
\nonumber
D_\mathrm{V}:=\{& x_1,x_2 \in \mathbb{R}:\\
\nonumber
&-(2^q+1-2^{-p}) \leq x_1,x_2 \leq 2^q+1-2^{-p}\land\\
\label{eq:FPValueDom}
& 2^{-p}\leq \vert x_1 - x_2 \vert\}.
\end{align}
The sign for a value \(x \in \mathbb{R}\) is realized by an additional bit that is either \(0\) for \(0 \leq x\) or \(1\) for \(x < 0\).
Violation of the value domain is called \textit{overflow} which may be caused during computations within the control algorithm.
For this example the overflow will be handled by wrapping since this type of implementations works without any additional logical tests.
The discretization is handled with a \textit{closest value} method that rounds the real value to the closest fixed-point value.
Not all digital processor provide overflow processing routines and some only support integer operations or even logical operation as well as the register shifts~\cite{NoArith}.
The aim of this work is to verify that a control algorithm does not produce overflow. The following example demonstrates that an overflow might lead to significant degradation of the controller performance.
\label{sec::DigiProcDef}

\subsection{Example}
\label{sec::Example}
Consider the following unstable plant:
\begin{align}
  \label{eq:SysStateSpace}
	A= \begin{bmatrix} 1 & 1 \\ 1& 0\end{bmatrix}, b=\begin{bmatrix}2\\0\end{bmatrix},c=\begin{bmatrix}0.5 \\ 0.5\end{bmatrix} ^{\top}, d=0
\end{align}
with the transfer function
\begin{equation*}
	G_{\mathrm{P}}(s) = \frac{s+1}{s^2-s-1}.
\end{equation*}
Furthermore assume that this plant has an additive uncertainty that is compensated with the weighting function \(W_1(s)=\frac{2s+1.1}{-10s^2-10.3s-5}\) and a multiplicative uncertainty that can be compensated with the weighting function \(W_2(s) = -3\).
To obtain a robust controller with respect to these uncertainties, we use the \(\mathcal{H}_{\infty}\) synthesis~\cite{Glover88} that results in the following transfer function:
\begin{equation*}
	G_{\mathrm{C}}(s)=\frac{2290 s^3 + 3775 s^2 + 2603 s + 707.8}{s^4 + 198.7 s^3 - 1332 s^2 - 1483 s - 767.9}.
\end{equation*}
\begin{figure}[!t]
  \center
  \vspace*{8px}
  \includegraphics[width=.4\textwidth]{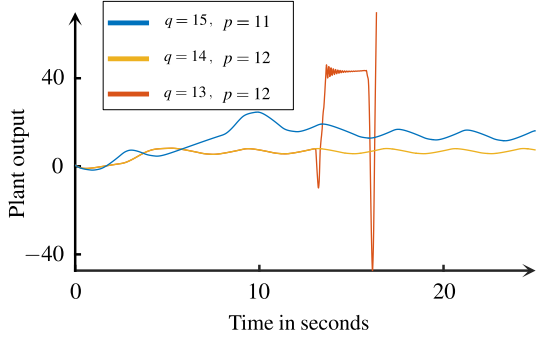}
  \caption{Simulation result of the system output with different integer and fraction part lengths.}
  \label{fig:Examples}
\end{figure}
We convert the controller transfer function \(G_{\mathrm{C}}(s)\) into a state space representation.
With respect to the implementation context we restrict the range of values for the matrices \(A \in D_{\mathrm{V}}^{4\times 4}, b\in D_{\mathrm{V}}^{4}, c\in D_{\mathrm{V}}^{1\times 4}\).
For the controller states \(z\in D_{\mathrm{V}}^4\) we specify the discrete-time implementation of the controller with system output \(y\) sampled at time \( h \cdot k, k \in \mathbb{N}\) and time-step size $h = 0.0001s$:
\begin{align}
	\nonumber
    z_{k+1} =&  z_{k} +h\colvec[0.9]{
      -1.03&   -0.5&    0&   0\\
      1 &  0 &   0 &  0\\
      0.0511 &   0.0216 &-513.9303 & 510.7020\\
      0.0314 &   0.0122 &-315.2434 & 316.2515
    }z_{k}+
    \\ 
    \nonumber
     &h\colvec[0.9]{
      5\\
      0\\
      -1023.4\\
      -632.5
    }y_k\\
    \label{eq:ControlSSRep}
  u_{k+1} =& \colvec[0.9]{
    0.0002 & 0.001 & -1.6189 & -1.0018
  }z_{k}
\end{align}%
We simulate the closed-loop system with different values for $p$ and $q$. 
Figure~\ref{fig:Examples} shows the result for an initial value for the system of \(x_{0}=[0,1]^{\top}\) and \(z_0 = 0\) for the controller.
The simulations only differ in the fixed-point specification.
The blue and yellow lines show successful stabilization.
They differ in the number of fraction bits but are similar in number of integer bits.
The orange and yellow lines have the same number of fraction bits, but reducing the the integer bits form \(14\) to \(13\) the controller to stops stabilizing the system after \(t=13s\).
This is caused by several overflow errors, which lead to the controller failure.

In the following sections we provide the approach to guarantee that a controller implementation will be overflow-free within the fixed-point arithmetic.
\vspace{-0.2cm}

\section{Proof Assistant System and Formal Defintions}
\label{sec::Prelim}
A partial goal of this work is to model the physical limitations of bit streams in \mct{Minlog}. Due to the functional nature of the modelling, \mct{Minlog} offers a solid set of tools for our purposes.
\vspace{-0.2cm}
\subsection{Introduction to \mct{Minlog}}
We introduce some preliminary concepts of the proof system \mct{Minlog}.
Roughly speaking, a proof assistant is a tool which is capable to address mathematical theory programmatically.
Within \mct{Minlog} we are able to define propositions \(P,Q\) to be arbitrary formulas and \(\rightarrow\) the logical implication.
With these symbols we can state
\begin{myitemize}
  \item \(P \rightarrow Q\), \(Q\) can be derived from \(P\)
  \item \(\forall \alpha .\; P(\alpha)\), the proposition \(P\) holds for all \(\alpha\).
\end{myitemize}

That way, we can define base types by constructor functions, further on called \textit{constructor}, that are either constant or recursive calls.

For example consider the natural numbers, that can be defined by $\mctm{zero}$ and a successor function \(\mctm{Succ}:= \mathbf{N} \Rightarrow \mathbf{N}\), mapping from the natural numbers into the natural numbers.
It is also possible to define types by other types. For instance the rational numbers, defined by numerator and denominator, may be represented by \(\mctm{RatConst}:=\mathbf{Z}\Rightarrow\mathbf{P}\Rightarrow\mathbf{Q}\).


With the type definitions and operators, proofs can be stated in the given logic and reasoned.
This is the setting a proof assistant offers.

Functions that are defined in \mct{Minlog} are called \textit{program constant}.
Since proofs and programs are equivalent, it is apparent that a program constant is proof of the correctness of the function~\cite{kennedy2011set}.
This forces the program constants to guarantee an output in the final type in a finite time.
This principle is called the \textit{notion of totality} and it ensures that the program will always yield correct results effectively.
In the present work, the totality of all program constants have been implemented.

To prove theorems with \mct{Minlog}, assumptions need to be provided which are relevant for the problem statement, e.g. by introducing propositions.
In \mct{Minlog} it is also possible to introduce the so-called inductively defined predicate constant (IDPC).
Informally speaking, these are assumptions specific to a variable. In following we separate the usual mathematical notation of numbers from the functional definitions within \textit{Minlog} by using bold letters rather than the double stricken.
For instance, the type realization for the rational numbers \(\mathbb{Q}\) is denoted by \(\mathbf{Q}\).

\subsection{Data representation types}
\label{sec::FPandSFP}
To model the behavior of the data in a finite storage processing system, we need to encapsulate the essential properties of this type.
Thus a representation for one and zero, a terminal symbol to limit the variable, and a terminal symbol marking an overflow.
We introduce the type \mct{fp} that we call a binary stream.
\begin{definition}[\mctm{Binary Stream}]
  The type \mct{fp} is defined by
  \begin{myitemize}
    \item \(\circ := \mctm{fp}\), the end of a bit stream, denoted by \(\circ\) ,
    \item \(\varnothing := \mctm{fp}\), the end of a bit stream with an overflow, denoted by \(\varnothing \),
    \item \(1 := \mctm{fp}\Rightarrow\mctm{fp}\), represents \(1\), and
    \item \(0 := \mctm{fp}\Rightarrow\mctm{fp}\), represents \(0\)
  \end{myitemize}
  and is denoted by the variable \(v\).
\end{definition}
\begin{remark}
  With these functions, we are capable to represent a generic bit-stream and mark whether it will return a correct value.
  A variable that has an overflow is called \textit{dirty}.
  Checking for the overflow is realized via the program constant \mct{FpNotOVL} which is defined such that \mct{FpNotOVL} returns false if the terminal symbol of the variable is \mct{OVL} or \(\top\) when the variable is clean.
\end{remark}

The main property of such bit-stream is the limitation of the length.
Program constants need to take this into account.
Thus, we define two versions for all program constants either limit the variable length, denoted with a superscript ``A'', or not.
Only limiting program constants can cause an overflow.
With the function \(\mctm{FpLength}:=fp\Rightarrow \mathbf{N}\), denoted as \(\langle v \rangle\) the length of a variable can be obtained.

We now present an interpretation of the data type \mctm{fp} that implements the fixed-point numbers.
The constructor functions themself represent the sign of the typed variable while the arguments represent the fraction and integer part of the fixed-point type.
\begin{definition}[\mctm{Signed Fixed Point Type}]\label{def:SfpDef}
  The signed-fixed point type \mct{sfp} is defined by
  \begin{align*}
    \mctm{FpPos} := \mctm{fp} \Rightarrow \mctm{fp}\Rightarrow \mctm{sfp}\\
    \mctm{FpNeg} := \mctm{fp} \Rightarrow \mctm{fp}\Rightarrow \mctm{sfp}
  \end{align*}
  and is denoted by the variable \(w\).
  The constructor \mct{FpPos} is represented with \(+(v_1,v_2)\) while \mct{FpNeg} is read as \(\sim(v_1,v_2)\), accordingly, where \(v_1\) is the fraction, and \(v_2\) integer.
\end{definition}
The multiplication in the \mctm{sfp} type is defined by a composure of limiting and nonlimiting program constants of the \mctm{fp} type.
\begin{definition}[\mct{SfpTimes}]\label{def:SfpTimes}
  The multiplication of the type \mctm{sfp} \(+(v_5,v_6) = +(v_1,v_2)\cdot +(v_3,v_4)\) is defined by
  \begin{align*}
    v_5=&\nu[\left(\left(v_1\cdot_{fp} v_4\right)+_{fp}\left(v_2\cdot_{fp} v_3\right)\right) +_{fp}\\
    &\iota\left(v_1 \cdot_{fp} v_3,\vert v_2\vert,\vert v_2\vert\right),\vert v_1\vert]\\
    v_6=&(v_2 \cdot_{fp}^{A} v_4 +_{fp}^{A}\\
    &\iota (\left(\left(v_1 \cdot_{fp} v_4\right)+_{fp}\left(v_2\cdot_{fp}v_3\right)\right),\vert v_1\vert,\vert v_2\vert))
  \end{align*}
  for fraction and integer part.
  The function \(\iota := \mctm{sfp}\Rightarrow\mathbf{N}\Rightarrow\mathbf{N}\Rightarrow \mctm{fp}\) selects a part of a bit-stream defined by the starting bit and amount of selected bits.
  \(\nu:= \mctm{fp}\Rightarrow \mathbf{N}\Rightarrow\mctm{fp}\) selects a given number of bits, but starting at the first bit.
\end{definition}
To connect the type with normal number definitions conversion functions are necessary.
Using \mct{Minlog} machinery we were able to prove the following technical theorems.
\begin{thm}[\mct{SfpToRatToSfpId}]\label{thm:SfpToRatToSfp}
  \begin{itshape}
    The conversion of a variable of the type \mctm{sfp} into the type of the rationals is without loss of accuracy. That is
    \begin{equation*}
      \forall w.\;\eta_{sfp}^{\mathbf{Q}}(\eta_{\mathbf{Q}}^{sfp}(w)) = w.
    \end{equation*}
  \end{itshape}
\end{thm}
\phantom{b}
\begin{thm}[\mct{RatToSfpToRatError}]\label{thm:RatToSfpToRat}
  \begin{itshape}
    The conversion of a variable \(a\) of the type \(\mathbf{Q}\) into a variable of the type \mctm{sfp} has, for the function \(\eta_{\mathbf{Q}}^{sfp}(a,p,q)\), an error bound of
    \begin{equation*}
      \forall a.\; \vert a\leq 2^q  \vert \rightarrow \vert \eta_{sfp}^{\mathbf{Q}}(\eta_{\mathbf{Q}}^{sfp}(a)) - a\vert \leq \frac{1}{2^{p+1}}
    \end{equation*}
    with \(p\) the amount of fraction bits and \(q\) the amount of integer bits.
  \end{itshape}
\end{thm}

The errors of the conversion in the operations are estimated as follows.
\phantom{b}
\begin{thm}[\mct{SfpPlusError}]\label{thm:SfpPlusErr}
  \begin{itshape}
    For the summation of two rational numbers \(a_1,a_2\) in the type \mctm{sfp} with \(p\) fraction bits and \(q\) integer bits the following property holds:
    \begin{align*}
      \forall a_1,a_2.\quad& \vert a_1+a_2 \vert \leq 2^q\rightarrow\\
      &\vert(\eta_{sfp}^{\mathbf{Q}}(\tra{\mathbf{Q}}{sfp}(a_1)+_{sfp}\tra{\mathbf{Q}}{sfp}(a_2))) \vert\leq 2\delta_{sfp}
    \end{align*}
    with \(\delta_{sfp} = 2^{-p}\).
  \end{itshape}
\end{thm}
\phantom{b}
\begin{proof}
 We use the the error estimates~(\ref{thm:SfpToRatToSfp}) and~(\ref{thm:RatToSfpToRat}).
  \begin{align*}
    \forall a_1&,a_2.\; \vert a_1+a_2 \vert \leq 2^q \rightarrow\\
    &\vert(\eta_{sfp}^{\mathbf{Q}}(\tra{\mathbf{Q}}{sfp}(a_1)+_{sfp}\tra{\mathbf{Q}}{sfp}(a_2)) - (a_1+a_2))\vert \leq\\
    &\vert a_1+\delta_{sfp}+a_2+\delta_{sfp} - (a_1+a_2)\vert =\vert 2\delta_{sfp}\vert
  \end{align*}
\end{proof}
\begin{thm}[\mct{SfpTimesError}]\label{thm:SfpTimesErr}
  \begin{itshape}
    For the multiplication of two rational numbers \(a_1,a_2\) in the type \mctm{sfp} with \(p\) fraction bits and \(q\) integer bits the following property holds:
    \begin{align}
      \nonumber
      \forall a_1,a_2.\quad & \vert a_1\cdot a_2 \vert \leq 2^q\rightarrow\\ \nonumber
      &\vert \tra{sfp}{\mathbf{Q}}(\tra{\mathbf{Q}}{sfp}(a_1)\cdot\tra{\mathbf{Q}}{sfo}(a_2)) - (a_1\cdot a_2)\vert \leq\\ 
      &\vert(a_1+a_2)\vert\delta_{sfp}+\delta_{sfp}^2
    \end{align}
    with \(\delta_{sfp} = 2^{-p}\).
  \end{itshape}
\end{thm}
\phantom{b}
\begin{proof}
  By Definitions~\ref{def:SfpTimes}, Theorem~\ref{thm:RatToSfpToRat} and under assumption \(\vert a_1\cdot a_2 \vert \leq 2^q\) we can write
  \begin{align*}
    &\vert \tra{sfp}{\mathbf{Q}}(\tra{\mathbf{Q}}{sfp}(a_1)\cdot\tra{\mathbf{Q}}{sfo}(a_2)) - (a_1\cdot a_2)\vert\\
    &\quad\leq \vert(a_1+\delta_{sfp})\cdot(a_2+\delta_{sfp}) - a_1\cdot a_2\vert\\
    &\quad=\vert(a_1+a_2)\vert\delta_{sfp}+\delta_{sfp}^2
  \end{align*}
\end{proof}
In the next section we derive the domain of the implementation that guarantees the overflow free computation of controller implementations.
\section{Main results}
\label{sec::MainRes}

Using the preliminary definitions from Section~\ref{sec::Prelim} the main result is provided in Theorem \ref{thm:ImpRelCor}.
We formally derive the property and proof that the computations within the given so-called reliable domain can be bounded.

We define a domain for the arguments of the implementation where we can assure that no overflow occurs, calling the results \textit{overflow free}, OF.
This domain is dependent on the sequence of operations of the implementation.
\begin{definition}[\mct{Sequenced Algorithm}]\label{def:SeqAl}
  A sequenced algorithm \(\Gamma(w)\) is a function where each binary base operation \(+,-,\cdot\), further depicted by \(\circ\), is assigned a value \(i \in \mathbb{N}\) that represents the place in the order of execution.
	We further on assume that \(\Gamma(w)\) is a function in fixed-point numbers \(D_{\mathrm{V}}\) that are defined by the number of integer bits and the number of fraction bits and \(w\in D_{\mathrm{V}}\).
\end{definition}
\phantom{b}
We now define the domain of the arguments where the results of the algorithm are correct.
\begin{definition}[\mct{Reliable Domain}]
  \label{def:RelDom}
  For a fixed-point type the implementation of the domain
  \begin{equation*}
    D_{\mathrm{R}} =\{ w : \Gamma_{\min} \leq \Gamma(w) \leq \Gamma_{\max}\}.
  \end{equation*}
  is called \textit{reliable domain}.
\end{definition}
\phantom{b}
For a specific fixed-point type definition, with \(q\) integer bits and \(p\) fraction bits, we can find, with Definition~\ref{def:RelDom} a minimum and a maximum value for \(\Gamma(w)\).
\begin{thm}[\mct{Bounded domain configuration}]
  \label{thm:ResDom}
  \begin{itshape}
  For a sequenced algorithm \(\Gamma(w)\) with $n$ operations \(\circ_i\) with the second argument given by \(\gamma_i \in D_{\mathrm{V}}\) and for a given \(w \in D_{\mathrm{V}}\) and \(p,q\in\mathbb{N}\) it holds that
  \begin{equation*}
    \forall w \in D_{\mathrm{V}},\;-2^q \leq \Gamma_{\min} \leq \Gamma(w)\leq\Gamma_{\max}\leq 2^q.
  \end{equation*}
  Furthermore, the domain \(D_{\mathrm{V}}\) is dependent on \(\Gamma(w)\), \(\Gamma_{\min}\) and \(\Gamma_{\max}\).
\end{itshape}
\end{thm}
\phantom{b}
\begin{proof}
  We proof by induction over \(n\).
We proceed by case distinction on arithmetical operations.The first case is for \((+)\):
  \begin{align*}
    &-2^q\leq \gamma_1 + w \leq 2^q\\
    &D_{\mathrm{V},min} = 2^q - \gamma_1 \leq w \leq 2^q -\gamma_1 = D_{\mathrm{V,max}}.
  \end{align*}
  The second case for the operation \((\cdot)\):
  \begin{align*}
    &-2^q\leq \gamma_1 \cdot w \leq 2^q\\
    &D_{\mathrm{V,min}} = 
     \frac{-2^q}{\gamma_1} \leq w \leq \frac{2^q}{\gamma_1} = D_{\mathrm{V,max}}.
  \end{align*}
  With the limitation of \(D_{\mathrm{V}}\) we obtain the maximum and minimum value for the implementation \(\Gamma_{\min} = \Gamma(D_{\mathrm{V,min}})\) and \(\Gamma_{\max} = \Gamma(D_{\mathrm{V,max}})\).
  
  In the induction step we again need to separate the tow cases.
  For \((+)\) the lower bound is
  \begin{align*}
    D_{\mathrm{V,min}}&\leq w + \Gamma(w) \leq w + \Gamma_{\min} \rightarrow D_{\mathrm{V,min}}-\Gamma_{\min}\leq w
  \end{align*}
  and for the upper bound
  {\small
  \begin{align*}
    w + \Gamma(w) \leq w + \Gamma_{\max} \leq D_{\mathrm{V,max}} \rightarrow w\leq D_{\mathrm{V,max}}-\Gamma_{\max}.
  \end{align*}
  }

  For \((\cdot)\) and a negative \(\Gamma_{\min}\) we obtain for the lower bound
  \vspace{-0.2cm}
  \begin{align*}
    D_{\mathrm{V,min}} \leq w \cdot \Gamma(w) \leq w \cdot \Gamma_{\min} \rightarrow  w\leq \frac{D_{\mathrm{V,min}}}{\Gamma_{\min}}.
  \end{align*}
  and for the upper bound
  \vspace{-0.2cm}
  \begin{align*}
    w \cdot \Gamma(w) \leq  w \cdot \Gamma_{\max}\leq D_{\mathrm{V,max}} \rightarrow \frac{D_{\mathrm{V,max}}}{\Gamma_{\max}} \leq w.
  \end{align*}
  Additionally the values \(\Gamma_{\min},\Gamma_{\max}\) can be influenced by the parameters.
  We follow the induction again, this time we assume operations without the argument \(w\).
  We have one case for \((+)\) since the case for \((\cdot)\) is trivial.
  \begin{align*}
    \Gamma_{\min}=-(2^q+1-2^{-p})\leq \gamma_1 \leq (2^q+1-2^{-p}) = \Gamma_{\max}
  \end{align*}

  For the induction step we have again two cases.
  For \((+)\) we have for the upper bound
  {\small
  \begin{align*}
    \Gamma_{\mathrm{min},i-1} \leq \gamma_i + \Gamma(w) \leq \gamma_i + \Gamma_{\min,i} \rightarrow \Gamma_{\min,i-1} - \gamma_i \leq \Gamma_{\min,i}.
  \end{align*}}
  For the lower bound we have
  {\small
  \begin{align*}
    \gamma_i + \Gamma(w) \leq \gamma_i + \Gamma_{\max,i} \leq \Gamma_{\mathrm{max},i-1} \rightarrow \Gamma_{\max,i} \leq \Gamma_{\mathrm{max},i-1} - \gamma_i.
  \end{align*}}
  For \((\cdot)\) we have and under assumption that \(0 \leq \vert \gamma_i \vert\) we have for the upper bound
  \vspace{-0.2cm}
  \begin{align*}
    \Gamma_{\mathrm{min},i-1} \leq \gamma_i \cdot \Gamma(w) \leq \gamma_i \cdot \Gamma_{\min,i} \rightarrow \frac{\Gamma_{\mathrm{min},i-1}}{\gamma_i}\leq \Gamma_{\min,i}.
  \end{align*}
  For the lower bound we have
  \begin{align*}
    \gamma_i \cdot \Gamma(w) \leq \gamma_i \cdot \Gamma_{\max,i} \leq \Gamma_{\mathrm{max},i-1} \rightarrow \Gamma_{\max,i} \leq \frac{\Gamma_{\mathrm{max},i-1}}{\gamma_i}
  \end{align*}

\end{proof}



\label{sec::Imp}
With the Definition~\ref{def:RelDom} and the Theorem~\ref{thm:ResDom} we can now formalize the type \mctm{imp}.
\begin{definition}[\mct{Implementation}]
  The type \mctm{imp} is defined by the constructor
  \begin{equation*}
    \mctm{imp}:=\mctm{sfp}\Rightarrow\mctm{sfp}\Rightarrow(\mctm{sfp}\Rightarrow\mctm{sfp})
  \end{equation*}
  The first two variables are the bounds of the reliable set \(D_{\mathrm{R}}\), followed by the specification of the sequenced algorithm \(\Gamma :=\mctm{sfp}\Rightarrow\mctm{sfp}\).
\end{definition}

We now further define the type with an IDPC that ensures correct processing of the given algorithm.
\begin{definition}[\mct{Reliable Implementation}]\label{def:ImpReliable}
  An implementation \(imp := w_1\Rightarrow w_2 \Rightarrow \Gamma(w)\) is called reliable if 
  \begin{equation*}
    \forall w.\, w_1 \leq w \leq w_2 \rightarrow \Gamma_{\mathrm{min}} \leq \Gamma(w) \leq \Gamma_{\mathrm{max}} \rightarrow \emptyset(\Gamma(w)).
  \end{equation*}
\end{definition}
\begin{remark}
  This ensures that the reliable domain is stated within the constructor and not an arbitrary set.
\end{remark}

We will now show the correctness of this type of implementation.
\begin{thm}[\mct{ImpRelCorr}]\label{thm:ImpRelCor}
  \begin{itshape}
    For a reliable implementation \(\Gamma(w)\), with the operation parameters \(\gamma_i\) and \(\gamma_m = \max{\vert\gamma_i\vert} \in D_{\mathrm{V}}\), the error to the rational valued polynomial \(g(a)\) is a bounded function of the argument \(a\) if \( \tra{fp}{\mathbf{Q}}(a) \in D_{\mathrm{R}}\). 
  \end{itshape}
\end{thm}
\begin{proof}
  The error of the function is defined by
  \begin{equation*}
    \delta = \left\vert g(a) - \tra{fp}{\mathbf{Q}}\left(\tra{\mathbf{Q}}{fp}\left(g(a)\right) \right) \right\vert.
  \end{equation*}
  We can estimate the error with Theorem~\ref{thm:SfpPlusErr} and~\ref{thm:SfpTimesErr} for a \(w \in D_{\mathrm{R}}\)  with \(\delta_c\) the conversion error and \(\delta_{t}\) the summation error.
  {\small
  \begin{align*}
    \delta &= \left\vert \sum_{i=0}^{n}\gamma_{i}a^i - \tra{fp}{\mathbf{Q}}\left(\tra{\mathbf{Q}}{fp}\left(\sum_{i=0}^{n}\gamma_{i}a^i\right)\right)\right\vert \\
    &\leq \left\vert \sum_{i=0}^{n}\gamma_{m}a^i - ((\gamma_{m} + \delta_c){(x+\delta_c)}^i + \delta_t)\right\vert \\
    &\leq \left\vert \sum_{i=0}^{n+1} \gamma_{m}\left(\sum_{k=1}^n \begin{pmatrix}n\\k\end{pmatrix}x^{n-k}\delta_c^{k}\right)+\delta_c{\left(a+\delta_c\right)}^n+\delta_t\right\vert \\
    &=\left\vert\frac{n^2+n}{2} \left(\delta_c a^n + (\gamma_{m}+\delta_c)\left(\sum_{k=1}^n \begin{pmatrix}n\\k\end{pmatrix}a^{n-k}\delta_c^{k} \right)+\delta_t\right)\right\vert.
  \end{align*}
  }
  This is a function of \(a\) and \(\Gamma\).
\end{proof}
\begin{remark}
  If the \(w \notin D_{\mathrm{R}}\) the implementation \(\Gamma(w)\) will have an overflow somewhere within the computation steps.
  This reduces the result of \(\Gamma(w)\) to a arbitrary value that might not stabilize the system anymore.
\end{remark}

\section{Case study}
\label{sec::CaseS}
To illustrate the formalization, we will show that the failing example from Section~\ref{sec::Problem} is not fulfilling Definition~\ref{def:ImpReliable}.
When we use the maximum controller state \(z_{max}\) during the stabilization process we can estimate the Equation~(\ref{eq:ControlSSRep}) with an upper bound:
\vspace{-0.2cm}
\begin{align*}
  u_{k+1} &\leq c_{\mathrm{C}}^{\top}(A_{\mathrm{C}}z_{max}+b_{\mathrm{C}}y(t))\\
  &=c_{\mathrm{C}}^{\top}A_{\mathrm{C}}z_{max}+(c_{\mathrm{C}}^{\top}b_{\mathrm{C}})y(t).
  \vspace{-0.2cm}
\end{align*}
\begin{table}
\vspace*{8px}
  \caption{The simulation parameters with the resulting bound of the reliable domain.}
  \label{tab:RunValues}
  \centering
  \begin{tabular}{ccccc}
          & \(p\) & \(q\) & \(D_{\mathrm{R,min}}\) & \(D_{\mathrm{r,max}}\) \\\hline
    \(1\) & \(11\) & \(15\) & \(-1.0352\) & \(27.5781\) \\
    \(2\) & \(12\) & \(14\) & \(-2.5435\) & \(11.7632\) \\
    \(3\) & \(12\) & \(13\) & \(1.0332\) & \(8.1863 \)
  \end{tabular}
\end{table}
\begin{figure}[t]
  \centering
  \includegraphics[width=.5\textwidth]{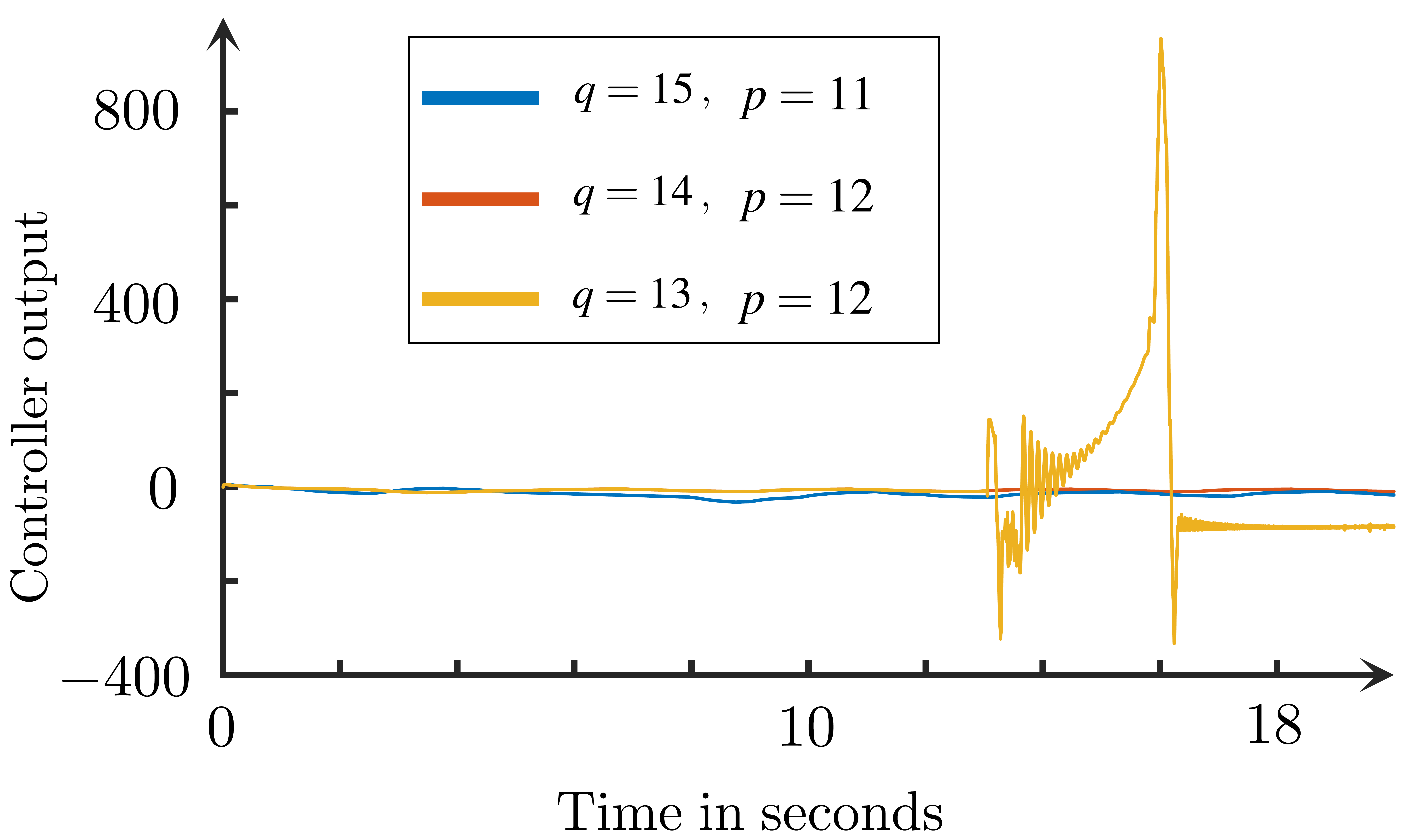}
  \caption{The output of the controller from Figure~\ref{fig:CtrOut} with their different fixed-point realizations.}
  \label{fig:CtrOut}
  \vspace{-0.6cm}
\end{figure}
We estimate these values for the different fixed-point type specifications and use them to obtain the bounds for the reliable domain \(D_{\mathrm{R}}\).
These are also dependent on the value domain and shown in Table~\ref{tab:RunValues} for the different specifications in Section~\ref{sec::Problem}.
We see that the first implementation contains the control target \(y=0\) and is capable of yielding correct results.
This is also shown in Figure~\ref{fig:CtrOut}.
For the yellow and red line the difference is that the control target is not within the reliable domain of the yellow simulation.
This lead to the destabilization of the feedback loop due to overflows.
What is interesting is that the controller performed reasonably well before the degradation occurred \(t=13s\).
It is believed that such effects may be overlooked when computations become more complex.

Thus, with this formalization of an implementation it becomes possible to check a priori whether the used specification for the fixed-point type is sufficient to achieve the control goals.
This way one may guarantee that the computations are overflow-free within the given domain and under the given assumptions, for all values of the domain.
In the case of the examples one may see that one additional bit for the integer part of the fixed-point type is sufficient to prevent overflows in the controller computation.
With the formalization this can be detected and the implementation can be adjusted to ensure that it is reliable.
\vspace{-0.1cm}
\section{Conclusion and Outlook}
\label{sec::Outl}
Even in case of robustly stabilizing controllers, the final software implementation of the control algorithms on digital processors can become unstable due to the overflow of the computations using fixed-point arithmetic.
In this work we have presented a formal method to verify that fixed-point representations on a processor are overflow free.
This was achieved by a formalization of the problem and by creating a formal type that incorporates the limitations of the digital processor.
For the derivations, the proof assistant system \mct{Minlog} was used.
The performance of the formally verified controllers was demonstrated in a case study.

One advantage of the presented approach is that it allows guaranteeing correctness of the mathematical derivation and of the software implementation simultaneously.
While there are methods for overflow handling it is more important to ensure that they can be avoided.
This then leads to smaller source code implementations and faster execution times.
Future work will consider more complex function evaluations as well as the predictive controllers.
\vspace{-0.2cm}
\bibliographystyle{abbrv}
\bibliography{bib/REf,bib/numerics,bib/control-theory,bib/form-ver-ctrl,bib/Examples}
\end{document}